\documentclass[10pt,conference]{IEEEtran}
\IEEEoverridecommandlockouts
\usepackage{cite}
\usepackage{amsmath,amssymb,amsfonts}
\usepackage{graphicx}
\usepackage{textcomp}
\usepackage{xcolor}
\def\BibTeX{{\rm B\kern-.05em{\sc i\kern-.025em b}\kern-.08em
    T\kern-.1667em\lower.7ex\hbox{E}\kern-.125emX}}

\usepackage{amsthm}
\newcommand{\R}{\mathbb{R}}

\newcommand{\ket}[1]{\left|#1\right\rangle}
\newcommand{\bra}[1]{\left\langle#1\right|}
\newcommand{\ketbra}[1]{| #1 \rangle \langle #1 |}
\usepackage{tikz}
\usetikzlibrary{arrows.meta, positioning}
\usepackage{graphicx}
\usepackage{algorithm}
\usepackage{algpseudocode}
\newtheorem*{remark}{Remark}

\usepackage{stfloats}
\usepackage[caption=false, labelfont=sf,textfont=sf]{subfig}
\begin{document}

\title{End-to-End Neural and Quantum Transcoding for Compressed Latent Representation under Channel Noise\\

\thanks{\IEEEauthorrefmark{1} These authors contributed equally to this work.}
}

\author{
\IEEEauthorblockN{Hyunho Cha\IEEEauthorrefmark{1}, Wonjung Kim\IEEEauthorrefmark{1}, and Jungwoo Lee}
\IEEEauthorblockA{\textit{NextQuantum and Department of Electrical and Computer Engineering, Seoul National University, Seoul, Republic of Korea}}
\IEEEauthorblockA{\{ovalavo, dnjswnd116, junglee\}@snu.ac.kr}
}

\maketitle


\begin{abstract}

Recent advancements in quantum computing highlight the need for efficient encoding of classical data into quantum states to ensure robust quantum information processing. Traditional encoding schemes often impose impractical requirements about the knowledge of quantum states and lack adaptability to noisy quantum channels and broader tasks. To address these limitations, we propose a novel end-to-end learnable quantum transcoding scheme explicitly optimized for compactness and robustness in noisy quantum communication scenarios. Our approach integrates neural network-based data compression with Cholesky decomposition-based quantum encoding and bypasses full density matrix reconstruction. Through normalized quantum observables, our method enables efficient tomography and achieves high reconstruction and classification performance even under extreme noise conditions.
\end{abstract}


\section{Introduction}

Quantum communication, a cornerstone of quantum information science, enables secure transmission, distributed quantum computing, and resource-efficient networking \cite{kimble2008quantum, bennett2014quantum}. Realizing these capabilities hinges on how information is represented and transmitted within quantum systems. Consequently, encoding classical data into quantum representations is foundational to effective quantum information processing and directly influences the performance of quantum-assisted tasks.

Traditional encoding schemes, such as qubit lattice \cite{venegas2003storing}, flexible representation of quantum images \cite{le2011flexible}, novel enhanced quantum representation \cite{zhang2013neqr}, Quantum Probability Image Encoding (QPIE)~\cite{yao2017quantum}, and so on, have been extensively studied. These early encoding schemes suffer from the limitation that only rough amplitude estimates can be extracted from few measurements~\cite{cavalieri2020quantum}.

More recently, hybrid quantum-classical machine learning frameworks have been employed for data compression \cite{romero2017quantum, pepper2019experimental, bischof2025hybrid}. To further reduce communication resources, quantum semantic communication using \(k\)-means clustering has emerged, which sends only the `semantic' features, rather than directly embedding classical data into quantum states~\cite{chehimi2024quantum}. Although it outperforms semantic-agnostic encoding schemes, \(k\)-means clustering might lose important semantic information when compression is aggressive. Also, it does not leverage prior knowledge of channel noise, which causes its performance to vary widely across different noise levels.
In fact, many existing methods do not adapt to error-prone quantum channels and are suboptimal in terms of qubit usage. Moreover, some reconstruction algorithms assume the complete knowledge of the quantum state after transmission \cite{wang2024quantum}, while in reality we can only approximate the state amplitudes with many measurements.

Addressing these limitations, we propose a novel learnable quantum transcoding scheme designed to optimize compactness of encoding and robustness against channel noise. In line with recent research, our scheme does not assume full reconstruction of the transmitted density matrix. Instead, it extracts a concise, informative feature vector through quantum observables, which are normalized with respect to the Hilbert-Schmidt norm to enable efficient tomography.
Furthermore, our framework incorporates awareness of both quantum channel characteristics and task requirements, adapting dynamically to varying noise conditions. Specifically, the contributions of our work include:
\begin{itemize}
\item A novel task- and noise-aware quantum transcoding scheme, combining neural data compression and Cholesky decomposition-based compact encoding.
\item Efficient bounded observable-based decoding that avoids the impracticality of full quantum state tomography.
\item Experimental validation on the MNIST dataset to confirm robust performance for both reconstruction and classification tasks across a broad range of noise levels.
\end{itemize}

This paper is structured as follows: Section~\ref{sec:background} reviews the essential concepts in quantum computing. Section~\ref{sec:method} details our proposed neural and quantum transcoding framework. Section~\ref{sec:experiments} presents experimental results. Finally, Section~\ref{sec:discussion} discusses the implications of our work and suggests potential directions for future research.

\section{Background}
\label{sec:background}

\subsection{Quantum Computing Fundamentals}

A quantum system is described by a state vector $\ket{\psi}$ residing in a Hilbert space $\mathcal{H}$. A Hilbert space is a complete vector space equipped with an inner product $\langle \cdot, \cdot \rangle$. In Dirac's bra-ket notation, the state $\ket{\psi}$ is a column vector in $\mathcal{H}$. The dual vector (bra) $\bra{\psi}$ is defined as the Hermitian conjugate of $\ket{\psi}$, and the inner product between two states is given by $\langle \phi | \psi \rangle$. This induces the norm $\|\ket{\psi}\| = \sqrt{\langle \psi|\psi \rangle}$.
For a qubit, the simplest quantum system, the state can be expressed as
\begin{align}
    \ket{\psi} = \alpha \ket{0} + \beta \ket{1}, \quad \alpha, \beta \in \mathbb{C}, \quad |\alpha|^2 + |\beta|^2 = 1.
\end{align}
Here, $\ket{0}$ and $\ket{1}$ form an orthonormal basis for the two-dimensional Hilbert space $\mathcal{H}_2$.
More generally, for a system with basis $\{\ket{i}\}_{i=1}^{D}$, any state can be written as
\begin{align}
    \ket{\psi} = \sum_{i=1}^{D} c_i \ket{i}, \quad c_i \in \mathbb{C}, \quad \sum_{i=1}^{D} |c_i|^2 = 1.
\end{align}

\begin{figure}[!t]  
\vspace{0.1in}
\centering

\includegraphics[width=0.44\textwidth]{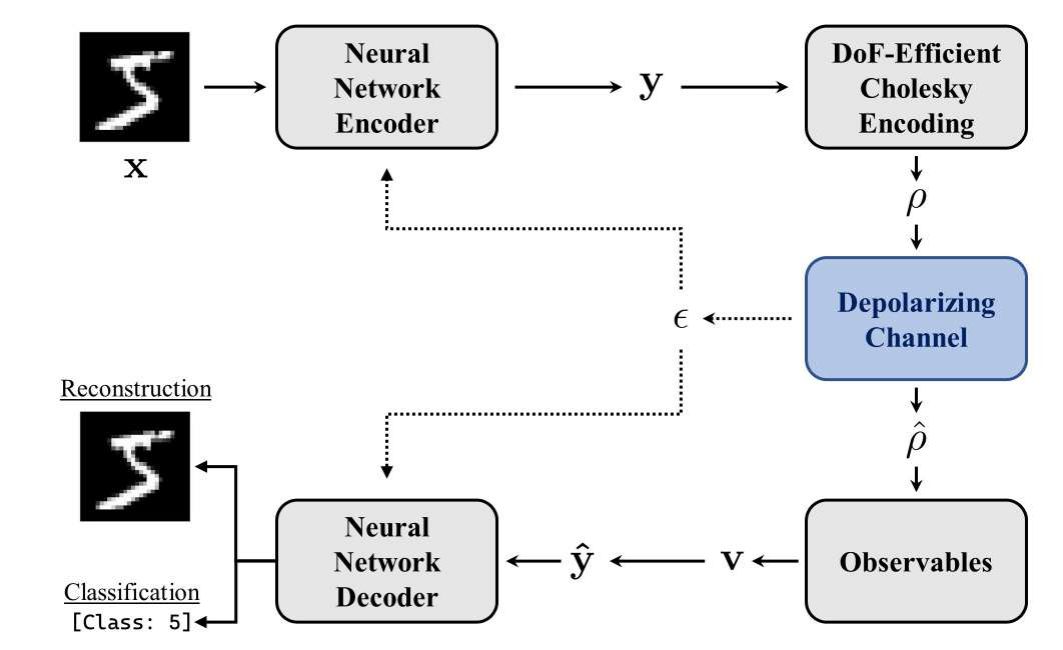}

\caption{Overview of the proposed learnable quantum encoding scheme. The neural network encoder and decoder both incorporate the noise parameter \( \epsilon \) to adapt to the quantum channel.}

\label{fig:flow}

\end{figure}  

\subsection{Density Matrix Formalism}

Quantum states may also be described using the density matrix formalism, which is particularly useful for representing mixed states or statistical ensembles of pure states. For a pure state $\ket{\psi}$, the density matrix is defined as \( 
\rho = \ketbra{\psi} \). For a mixed state, where the system is in state $\ket{\psi_i}$ with probability $p_i$, the density matrix is
\begin{align}
    \rho = \sum_i p_i\, \ket{\psi_i}\bra{\psi_i}, \quad p_i \geq 0, \quad \sum_i p_i = 1.
\end{align}
The density matrix $\rho$ satisfies the following properties
\begin{itemize}
\item \textbf{Hermiticity:} $\rho = \rho^\dagger$.
\item \textbf{Positivity:} $\langle \phi|\rho|\phi\rangle \geq 0$ for all $\ket{\phi} \in \mathcal{H}$.
\item \textbf{Normalization:} $\operatorname{tr}(\rho) = 1$.
\end{itemize}

\subsection{Observables and Measurements}

Observables correspond to measurable physical quantities and are represented by Hermitian operators \(O\). The spectral decomposition of $O$ is given by \( O = \sum_k o_k \Pi_k \), where $o_k$ are the eigenvalues and $\Pi_k = \ket{o_k}\bra{o_k}$ are the associated projection operators. When measuring an observable on a state $\rho$, the probability of obtaining the eigenvalue $o_k$ is computed by \( p(o_k) = \operatorname{Tr}(\rho \Pi_k) \). Therefore, the expectation value of the observable is
\begin{align}
\langle O \rangle = \sum_k o_k p(o_k) = \operatorname{Tr}(\rho O).
\end{align}
Quantum machine learning employs the expectation value \( \langle O \rangle = \operatorname{Tr}(\rho O) \) as a crucial bridge between quantum and classical computation \cite{huang2021power, cerezo2022challenges}. In variational algorithms, for example, this quantity defines the cost function that guides the optimization of parameterized quantum circuits \cite{cerezo2021variational}. Furthermore, when quantum states encode classical data, the resulting expectation values serve as informative features for subsequent classical processing, enabling efficient hybrid models that leverage both quantum advantages and classical robustness.

\section{Proposed Framework}
\label{sec:method}

As illustrated in Fig.~\ref{fig:flow}, the proposed framework consists of an outer neural network for semantic compression and an inner quantum layer for state encoding and transmission. This design enables efficient, task- and noise-aware mapping from classical inputs to quantum states. The following subsections describe the neural compression module and the quantum transcoding and decoding components in detail.

\subsection{Transformer-Based Encoder–Decoder Architecture with Latent Refinement}
\begin{figure}[!h]
    \centering
    \includegraphics[width=0.44\textwidth]{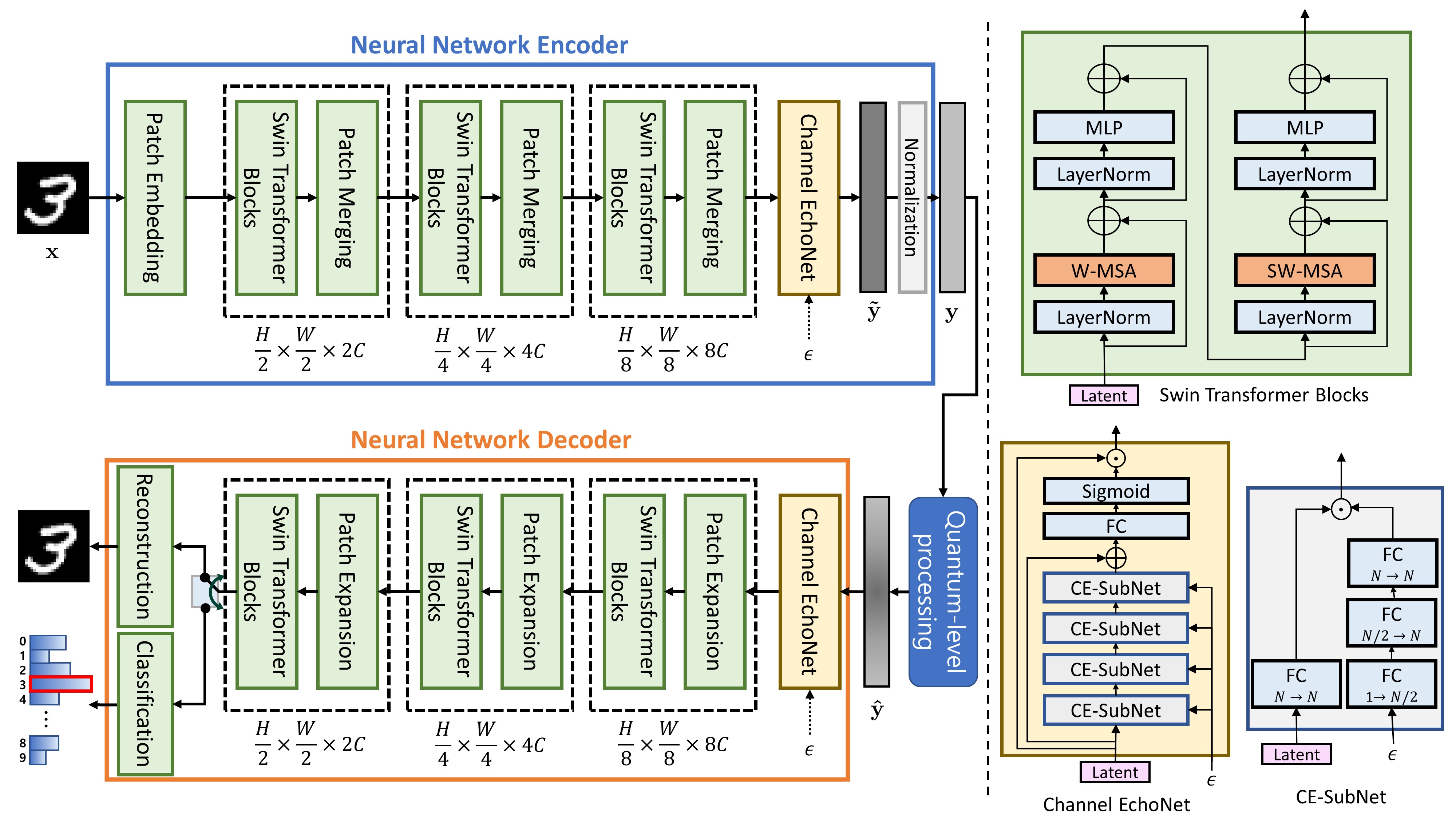}
    \caption{Encoder–decoder pipeline for neural compression of input images into latent representations.}
    \label{fig:swin_transformer}
\end{figure}

As illustrated in Fig.~\ref{fig:swin_transformer}, we propose a unified neural network architecture for joint image reconstruction and classification. The model follows an encoder–decoder paradigm, comprising a neural encoder, a neural decoder, and an intermediate module termed \emph{Channel EchoNet} designed to enhance latent representations against noisy channel. Both encoder and decoder adopt a hierarchical structure based on Swin Transformer blocks\cite{liu2021swin}, which efficiently model local and global dependencies via window-based and shifted window-based self-attention.

The encoder receives an input grayscale image $\mathbf{x} \in \mathbb{R}^{H \times W \times 1}$, where $H$ and $W$ denote the height and width of the image, respectively, and first partitions it into non-overlapping patches. These patches are projected into an embedding space and subsequently processed by a sequence of Swin Transformer blocks interleaved with patch merging layers. Each Swin Transformer block consists of window-based multi-head self-attention (W-MSA), shifted window-based attention (SW-MSA), layer normalization, and a multi-layer perceptron (MLP). Patch merging progressively reduces the spatial resolution while increasing the channel dimensionality, resulting in a compact multi-scale latent representation.

To further refine the latent representation, we introduce the Channel EchoNet, which modulates channel-wise activations through a sequence of lightweight sub-networks, referred to as CE-SubNets. Each CE-SubNet performs a non-linear transformation that expands and contracts the channel dimensionality, while residual connections are employed to preserve intermediate features and stabilize the training process.

After this refinement process, we obtain a latent representation denoted by $\mathbf{\tilde{y}}$, which is subsequently projected onto the unit $(N\!-\!1)$-dimensional hypersphere, yielding a normalized latent representation $\mathbf{y} \in S^{N-1}$, where $S^{N-1}$ represents the unit sphere in $\mathbb{R}^N$. This normalized representation serves as the input to the subsequent quantum transcoding stage, where it is mapped to a quantum state, transmitted through a noisy channel, and later recovered as a corrupted latent vector $\hat{\mathbf{y}}$. The details of this quantum processing pipeline will be presented in the next subsection.

The decoder mirrors the encoder structure, starting from the refined latent representation. Patch expansion layers and Swin Transformer blocks are applied to progressively restore spatial resolution. The output features are then passed to two heads: a reconstruction head that estimates the original image and a classification head that produces categorical logits. The proposed design facilitates the learning of semantically rich and robust representations, achieving effective performance across both classification and reconstruction tasks.

\subsection{Compact Quantum Transcoding from Cholesky Decomposition}

\begin{figure}
    \centering
    \includegraphics[width=0.9\linewidth]{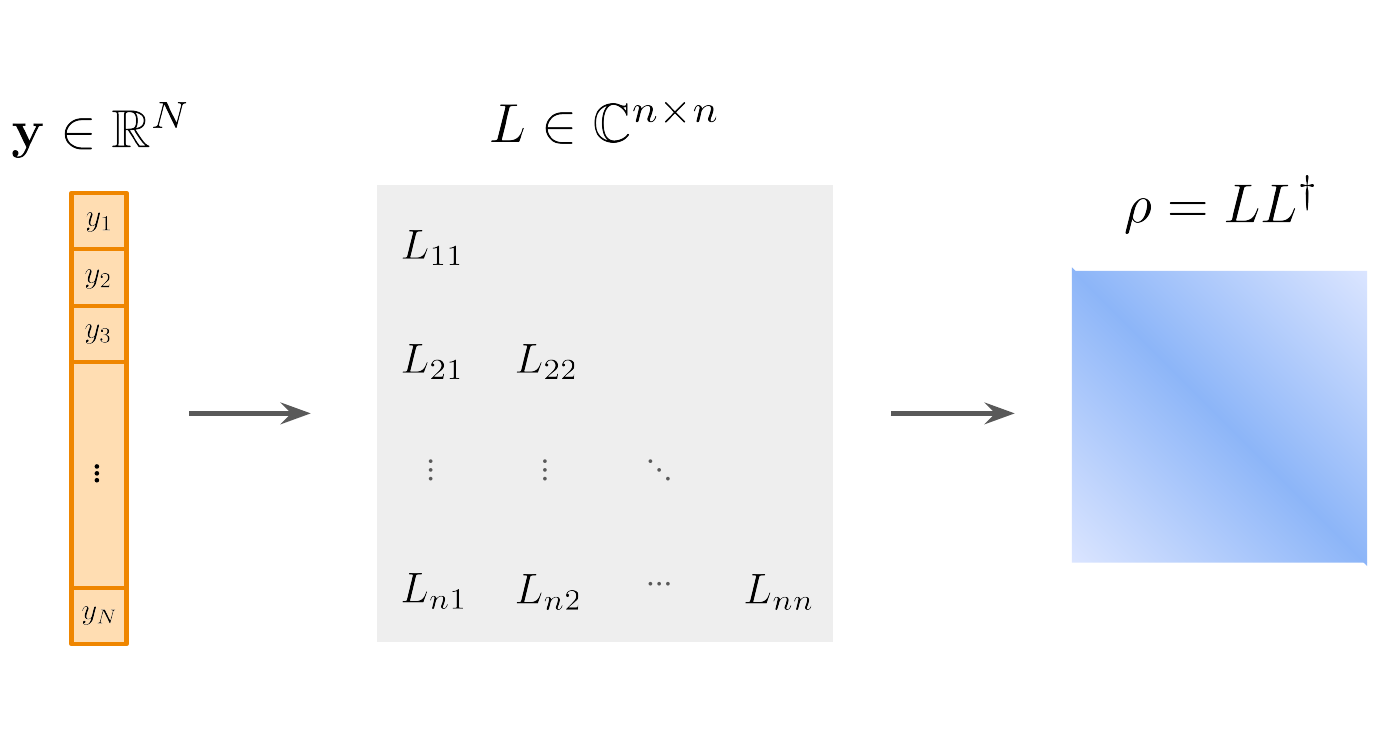}
    \caption{The proposed DoF-Efficient Cholesky Encoding. Components of \(\mathbf{y}\in\R^N\) are mapped onto the entries of the lower triangular matrix \(L\), where diagonal elements are real and off-diagonal elements are complex. For example, \(L_{11}=y_1\), \(L_{22}=y_2\), \(L_{21}=y_3+y_4i\), and so on.}
    \label{fig:dm_encoding}
\end{figure}

To establish a mapping $\mathbf{y}\mapsto\rho$, we can construct a complex lower triangular matrix (with real diagonal entries) \( L \in \mathbb{C}^{n \times n} \), where the entries of $L$ are parameterized by the elements of $\mathbf{y}$. The number of real parameters in \(L\) matches the degrees of freedom required by the density matrix, which is \( n^2 \).
We determine the smallest integer $n\in\mathbb{N}$ such that \( n^2 \geq N \), i.e., \( n = \lceil \sqrt{N} \rceil \). It defines the effective dimension of the Hilbert space $\mathcal{H}$ used in our encoding. Our goal is to construct a density matrix $\rho \in \mathcal{B}(\mathcal{H})$ (the space of bounded operators on $\mathcal{H}$) that faithfully represents the compressed feature vector.
Since $\rho$ has $n^2 - 1$ degrees of freedom and $\mathbf{y}$ has $N - 1$ degrees of freedom, the inequality \( n^2 - 1 \ge N - 1 \) ensures that there exists a mapping $\mathbf{y} \mapsto \rho$ such that no degree of freedom is lost in the process.
A key property of positive definite matrices is that they admit a unique Cholesky decomposition. Hence, if we set \( \rho = LL^\dagger \) and ensure that the diagonal entries of $L$ are strictly positive, then $\rho$ is guaranteed to be positive definite and \( \mathbf{y} \mapsto \rho \) is invertible \cite{trefethen2022numerical}. In practice, even if $\mathbf{y}$ contains both positive and negative elements (which may lead to some loss of sign information in \( \rho \mapsto \mathbf{y} \)), empirical evaluations show that the overall performance of the encoding is not compromised.
Note that
\begin{align}
    \text{tr}(\rho) = \text{tr}(L L^\dagger) = \|L\|_\text{F}^2 = \|\mathbf{y}\|_2^2 = 1,
\end{align}
where $\|\cdot\|_\text{F}$ denotes the Frobenius norm. Therefore, $\rho$ automatically becomes a valid density matrix. This encoding strategy is illustrated in Fig.~\ref{fig:dm_encoding}.

Mapping \(y\) to \(\rho\) can also be approached through the Bloch vector parameterization framework. However, as we shall see, this parameterization is not well-suited to our task. A generalized Bloch vector has been introduced and discussed in various works \cite{kimura2003bloch, kimura2005bloch, bengtsson2017geometry}.
To explicitly construct this mapping, one begins by defining a set of $n^2$ projection operators $\pi_{jk} = | j \rangle \langle k |$, where $j, k \in \{ 0, 1, \cdots , n - 1 \}$ denote computational basis states.

Next, define three types of operators and combine them:

\begin{subequations}
\begin{IEEEeqnarray}{RCL}
    \alpha_j & \equiv & \sqrt{\frac{2}{j(j + 1)}} \left( 
    \sum_{k = 0}^{j} \pi_{kk} - (j + 1) \pi_{j + 1, j + 1} \right), \\
    \beta_{jk} & \equiv & \pi_{jk} + \pi_{kj}, \quad j < k, \\
    \gamma_{jk} & \equiv & i(\pi_{jk} - \pi_{kj}), \quad j < k, \\
    \{ \lambda_i \} & \equiv & \{ \alpha_j \} \cup \{ \beta_{jk} \} \cup \{ \gamma_{jk} \}.
\IEEEeqnarraynumspace
\end{IEEEeqnarray}
\end{subequations}
The set $\{ \lambda_i \}$ contains $n^2 - 1$ operators satisfying the relations
\begin{eqnarray}
\label{eq:lambda_trace}
\text{tr}(\lambda_i) = 0 \quad \text{and} \quad \text{tr}(\lambda_i \lambda_j) = 2\delta_{ij}.
\end{eqnarray}

The fidelity between quantum states can be represented using their corresponding generalized Bloch vectors. Specifically, for an $n$-dimensional state $\rho$, its generalized Bloch vector $\mathbf{r}^{(\rho)} \in \mathbb{R}^{n^2 - 1}$ is defined via the decomposition
\begin{align}
\label{eq:bloch_definition}
\rho = \frac{1}{n} \left( I + \frac{n(n - 1)}{2} \lambda \cdot \mathbf{r}^{(\rho)} \right),
\end{align}
where $|| \mathbf{r}^{(\rho)} || \le 1$. Using (\ref{eq:lambda_trace}) and (\ref{eq:bloch_definition}), the purity of \( \rho \) can be expressed as \( \text{tr}(\rho^2) = \frac{1 + (n - 1) \| \mathbf{r}^{(\rho)} \|^2}{n} \). Thus, a state \(\rho\) is pure if and only if \( \| \mathbf{r}^{(\rho)} \|^2 = 1 \). Consequently, interpreting a unit vector \( 
\Tilde{y} \in \mathbb{R}^{n^2 - 1} \) as \( \mathbf{r}^{(\rho)} \) inherently restricts the representation to pure states.
Although it is possible to make \( \| \Tilde{y} \| < 1 \), a more critical issue remains. That is, for \( n > 2 \), not all elements of the unit ball centered at the origin are valid Bloch vectors. That is, the encoder output \( \Tilde{y} \) is not guaranteed to produce a physically realizable density matrix.

\subsection{Noisy Quantum Channel and Observable Readout}

After obtaining the density matrix $\rho$, the state is transmitted through a noisy quantum channel. In this work, we model the noise using a depolarizing channel \cite{nielsen2010quantum} defined as
\begin{align}
    \mathcal{E}_\epsilon(\rho) = (1-\epsilon)\rho + \frac{\epsilon}{n} I,
\end{align}
where $\epsilon\in[0,1]$ represents the noise parameter and $I$ is the identity operator on $\mathcal{H}$.
To extract features for subsequent processing, we define a set of $K$ parameterized observables $\{O_i\}_{i=1}^{\smash{K}}$, each satisfying \( O_i = O_i^{\smash{\dagger}} \) and \( \text{tr} \smash{(O_i^2)} = 1 \). These observables, which are standard in quantum machine learning, yield the feature vector
\begin{align}
    \mathbf{v} = \begin{bmatrix} \operatorname{tr}\left(\mathcal{E}_\epsilon(\rho) O_1\right) \\ \operatorname{tr}\left(\mathcal{E}_\epsilon(\rho) O_2\right) \\ \vdots \\ \operatorname{tr}\left(\mathcal{E}_\epsilon(\rho) O_K\right) \end{bmatrix} \in \mathbb{R}^K.
\end{align}
This vector of expectation values is used to reconstruct the latent representation as \( \hat{\mathbf{y}} = f_{\text{proj}}(\mathbf{v}) \), where \( f_{\text{proj}}(\cdot) \) is a learnable linear projection network for matching the dimension with \( \mathbf{y} \). Finally, \( \hat{\mathbf{y}} \) is used to decode and reconstruct the original image or to perform other tasks. Importantly, both the encoder and decoder networks are explicitly designed to be noise-aware. In the specific case of the depolarizing channel, which is the primary noise model considered in this work, the noise parameter \( \epsilon \) is directly fed into both the encoder and decoder. This enables the networks to adapt their behavior based on the channel condition, leading to more robust performance under varying noise levels.

\begin{remark}
Normalizing the Hilbert-Schmidt norm of the observables, i.e., \( \text{tr} (O_i^2) = 1 \), ensures that all \( K \) expectation values in the feature vector \( \mathbf{v} \) can be efficiently estimated via random Clifford measurements \cite[Theorem~2]{huang2020predicting}. Specifically, the \( K \) linear functions
\begin{align}
    \operatorname{tr}\left(\mathcal{E}_\epsilon(\rho) O_1\right), \, \ldots \, , \, \operatorname{tr}\left(\mathcal{E}_\epsilon(\rho) O_K\right)
\end{align}
can be simultaneously estimated up to additive error \( \varepsilon \) with probability \( \ge 1 - \delta \) using
\begin{align}
    O \left( \frac{1}{\varepsilon^2} \log \left( 
    \frac{K}{\delta} \right) \right)
\end{align}
copies of \( \rho \). Although the observables \( O_i \) are classically optimized and might not individually correspond to physically simple measurement settings, classical shadows circumvent this challenge by employing observable-independent measurements.
\end{remark}

\section{Experimental Results}
\label{sec:experiments}

This section presents a performance evaluation of the proposed framework over a depolarizing channel. A multi-task learning approach is adopted to jointly perform reconstruction and classification. The influence of the density matrix dimension \(n\) and the number of observables \(K\) is systematically analyzed under varying levels of channel noise. Experimental results show improved reconstruction robustness over the QPIE baseline and demonstrate reliable classification performance under channel noise.

\subsection{Experimental Setting}

We conduct all experiments on the MNIST dataset, which consists of grayscale images of handwritten digits with 10 classes. Each image is resized to \(32 \times 32\) pixels to ensure compatibility with the input size required by our network. The proposed model is trained and evaluated on two tasks—reconstruction and classification—to capture different aspects of its performance.

In the reconstruction task, the model is trained to minimize the mean squared error (MSE) loss, which is defined as
\begin{align}
\mathcal{L}_{\mathrm{MSE}} &= \frac{1}{N} \sum_{i=1}^{N} \| \mathbf{x}_i - \hat{\mathbf{x}}_i \|_2^2,
\end{align}
where \(\mathbf{x}_i\) and \(\hat{\mathbf{x}}_i\) are the original and reconstructed image vectors, respectively, and \(N\) is the number of samples. To evaluate reconstruction quality, we use Peak Signal-to-Noise Ratio (PSNR) and Structural Similarity Index Measure (SSIM). PSNR is defined as
\begin{align}
\mathrm{PSNR} &= 10 \cdot \log_{10} \left( \frac{L^2}{\mathrm{MSE}} \right),
\end{align}
where \(L\) is the maximum possible pixel value (e.g., 255 for 8-bit images). SSIM measures perceptual similarity and is computed as
\begin{align}
\mathrm{SSIM}(I_1, I_2) &= \frac{(2\mu_{I_1} \mu_{I_2} + C_1)(2\sigma_{I_1 I_2} + C_2)}{(\mu_{I_1}^2 + \mu_{I_2}^2 + C_1)(\sigma_{I_1}^2 + \sigma_{I_2}^2 + C_2)},
\end{align}
where \(\mu_{I_1}, \mu_{I_2}\) are the means, \(\sigma_{I_1}^2, \sigma_{I_2}^2\) are the variances, and \(\sigma_{I_1 I_2}\) is the covariance of images \(I_1\) and \(I_2\). \(C_1\) and \(C_2\) are constants to avoid instability.

In the classification task, the model is trained using the cross-entropy loss:
\begin{align}
\mathcal{L}_{\mathrm{CE}} &= - \sum_{i=1}^{N} \sum_{c=1}^{C} y_{i,c} \log \hat{y}_{i,c},
\end{align}
where \(y_{i,c}\) is the one-hot ground-truth label and \(\hat{y}_{i,c}\) is the predicted probability for class \(c\). The performance is evaluated using Top-1 accuracy, which represents the proportion of test samples where the predicted class with the highest probability matches the true label.

For the network backbone, we adopt a Swin Transformer configured with a patch size of 4 and a shifted window size of 2, adjusted to accommodate the small spatial resolution of the MNIST dataset. Furthermore, the main experimental variables under investigation are \( n \) —the dimension of the density matrix— and \( K \) —the number of observables. These variables are varied systematically to analyze their impact on both reconstruction and classification performance.
Additionally, the AdamW optimizer\cite{adamw} is employed for training, with a constant learning rate of $1.0 \times 10^{-4}$ and a total of 50 training epochs.

\subsection{Simulation Results in Reconstruction Task}

For the image reconstruction task, we adopt QPIE as a baseline, which is widely used in classical image encoding scenarios. To compare with our method, we evaluate the reconstruction performance under varying channel noise levels, using both PSNR and SSIM as evaluation metrics.

\begin{figure}[!h]
    \centering
    \includegraphics[width=0.33\textwidth]{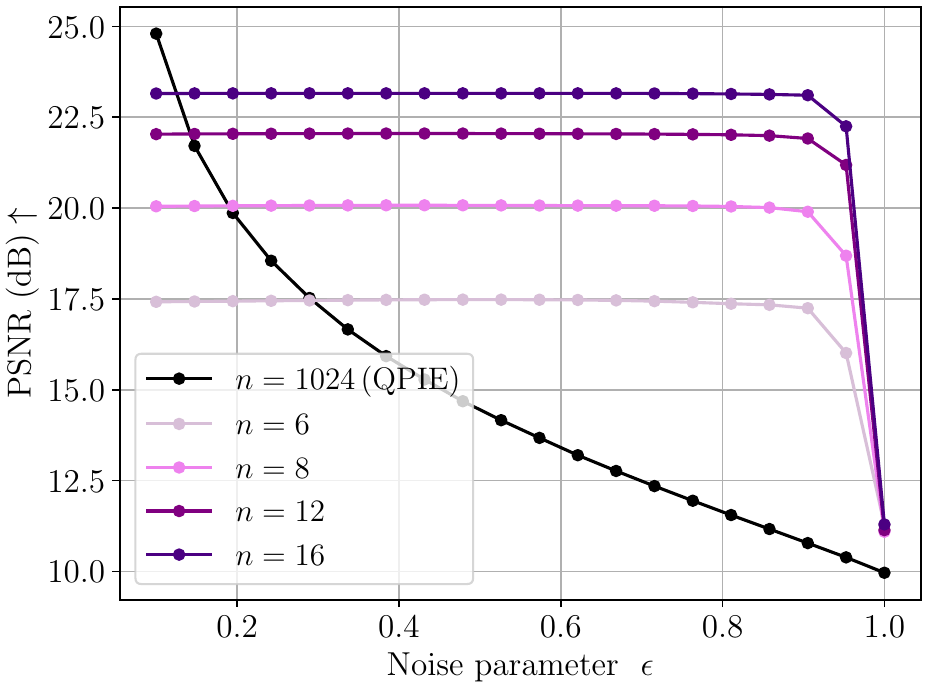}
    \caption{PSNR comparison across different dimensions of the density matrix with \(K = 10\). Higher PSNR indicates better reconstruction quality. Note that the case \(\epsilon = 0\) is omitted due to its theoretical divergence.}
    \label{fig:psnr_f}
\end{figure}
\begin{figure}[!h]
    \centering
    \includegraphics[width=0.33\textwidth]{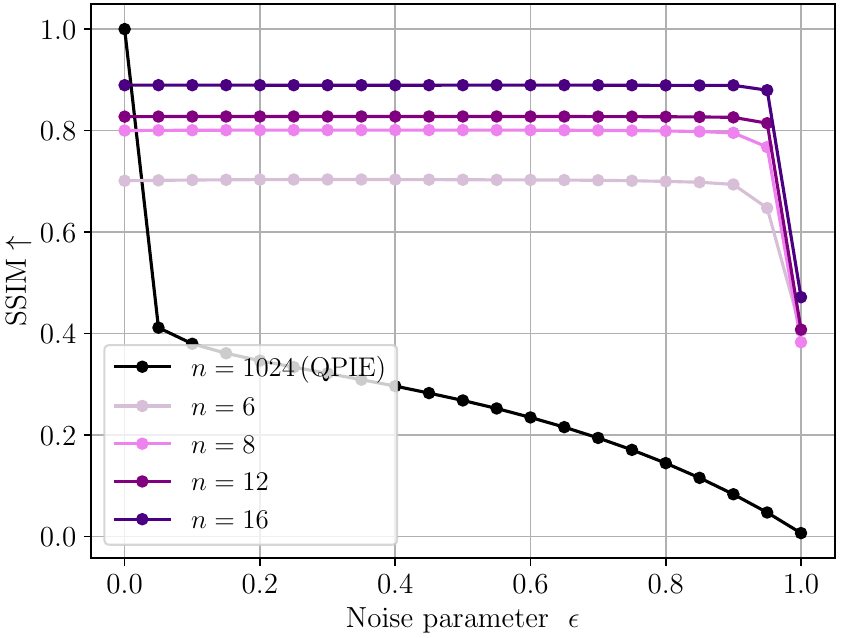}
    \caption{SSIM comparison across different dimensions of the density matrix with \(K = 10\). Higher SSIM indicates better structural similarity.}
    \label{fig:ssim_f}
\end{figure}

As shown in Figures~\ref{fig:psnr_f} and~\ref{fig:ssim_f}, the proposed method consistently outperforms the QPIE baseline across a wide range of compression ratios and noise levels. Notably, it maintains stable performance even under large values of \(\epsilon\), indicating strong robustness in highly noisy environments. In contrast, QPIE suffers from significant performance degradation as noise increases. Furthermore, the reconstruction quality improves with increasing values of \(n\), suggesting a trade-off between compression efficiency and reconstruction fidelity. These results demonstrate the effectiveness and noise-resilience of our method under diverse channel conditions.

\begin{figure}[!h]
    \centering
    \vspace{0.0cm}
    \subfloat[\tiny PSNR]{
        \includegraphics[width=0.5\linewidth]{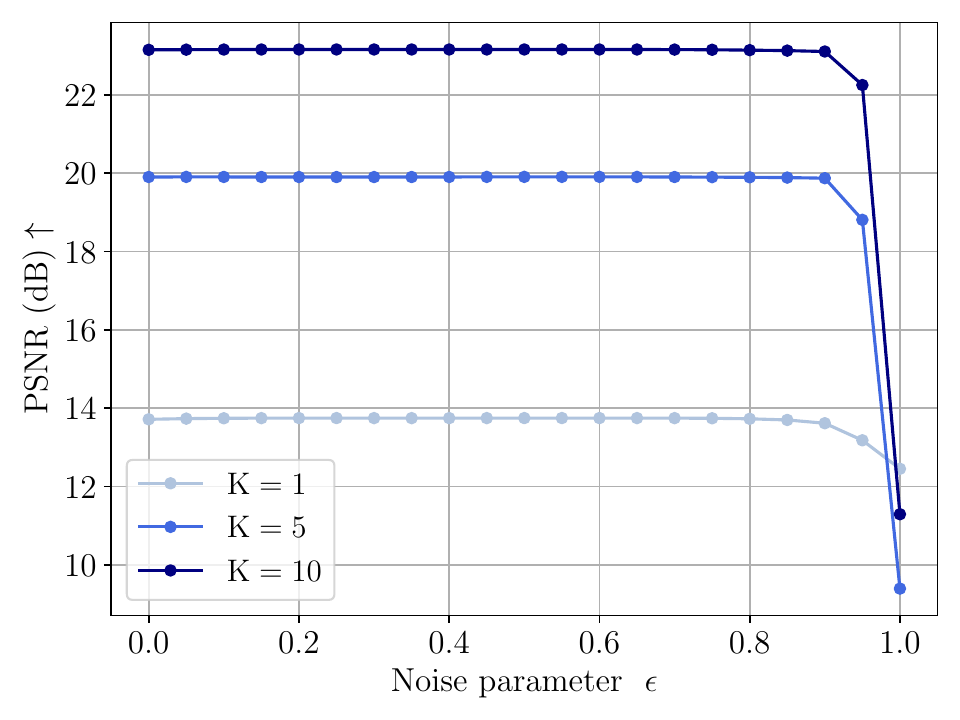}
    }
    \subfloat[\tiny SSIM]{
        \includegraphics[width=0.5\linewidth]{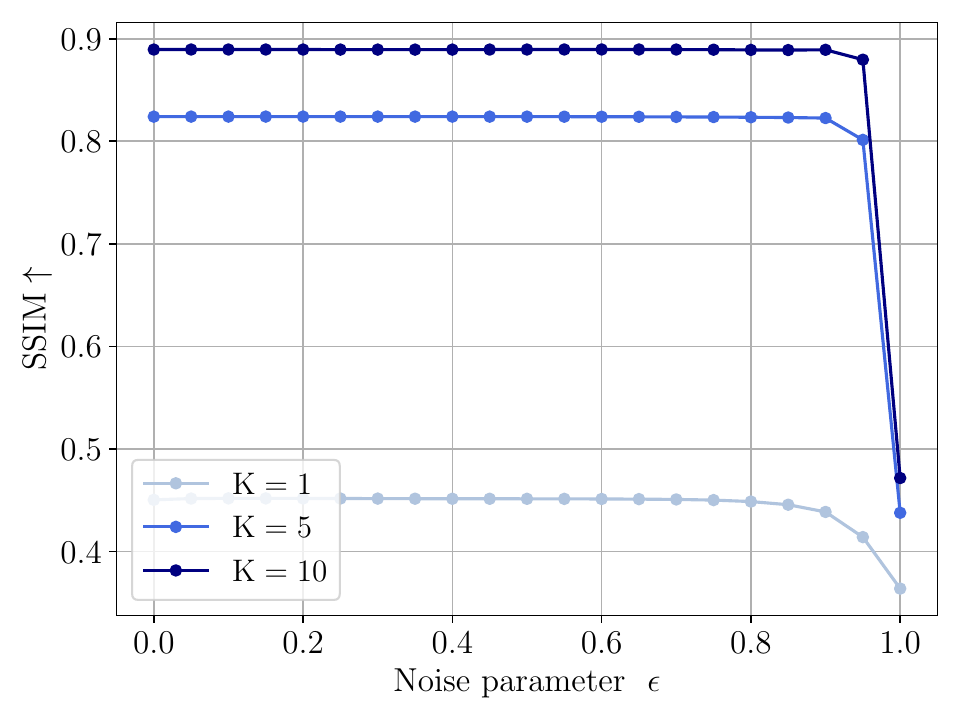}
    }
    \caption{Reconstruction performance comparison against different number of observables with $n=32$.}
    
    \label{fig:ssim_k}
\end{figure}

As shown in Fig. \ref{fig:ssim_k}, the results confirm the noise robustness of the proposed method across different \(\epsilon\) values. When \(K = 1\), the method produces reliable outputs from a single observation. Increasing \(K\) further improves performance, demonstrating the benefit of using multiple observations to enhance overall effectiveness.

\begin{figure}[!h]
    \centering
    \includegraphics[width=0.37\textwidth]{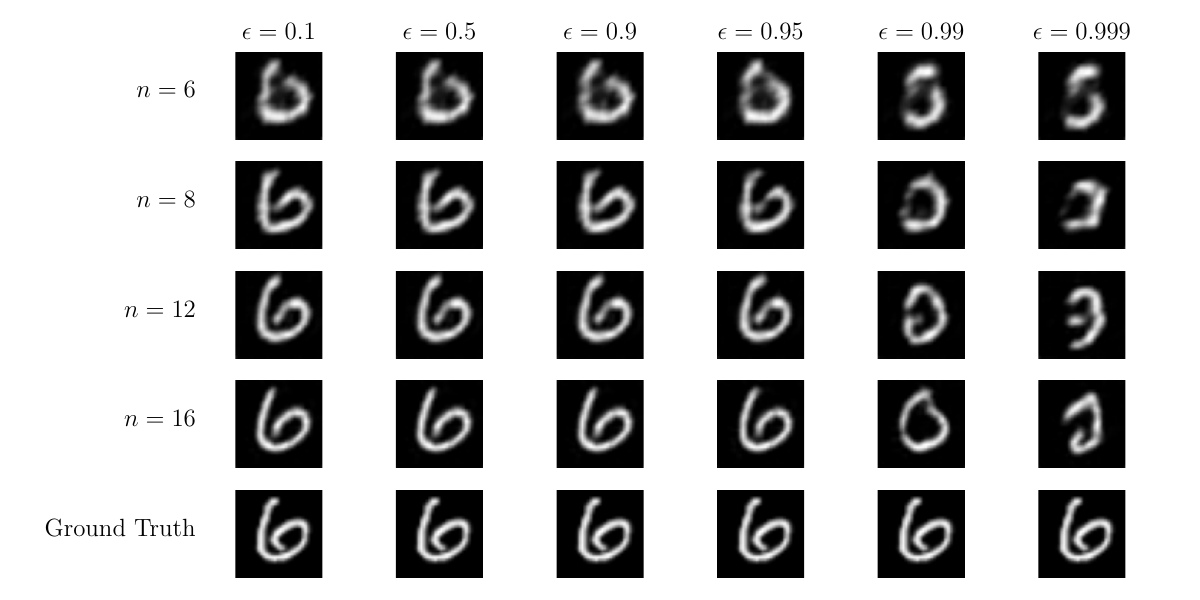}
    \caption{Image reconstruction example against different dimensions of the density matrix in $K=10$.}
    \label{fig:rec_f}
\end{figure}

\begin{figure}[!h]
    \centering
    \includegraphics[width=0.37\textwidth]{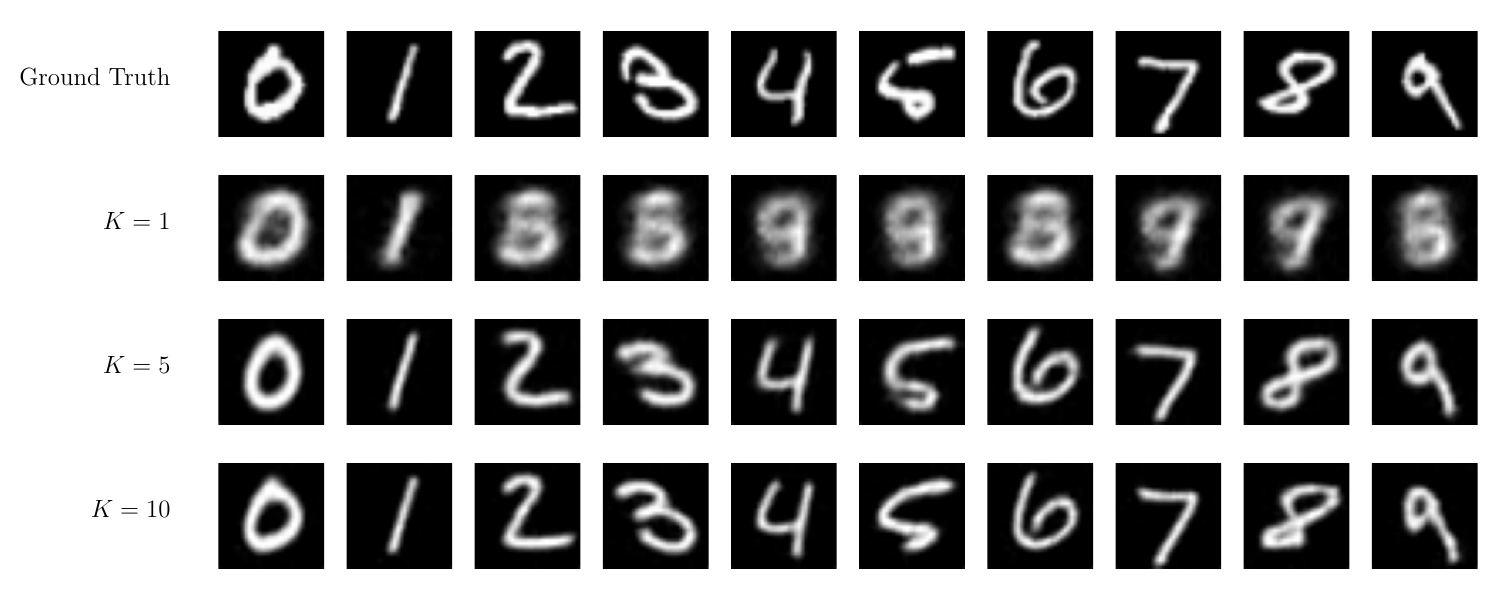}
    \caption{Image reconstruction example against different number of observables in $\epsilon=0.9$.}
    \label{fig:rec_k}
\end{figure}

Fig.~\ref{fig:rec_f} and Fig. \ref{fig:rec_k} illustrate reconstruction examples of the MNIST dataset for different values of \(n\) and \(K\). As both \(n\) and \(K\) increase, the reconstructed images exhibit reduced blurring and improved clarity, confirming the benefit of using a larger number of observables for better reconstruction quality. While this trend holds across all noise levels, it is particularly noteworthy that the method generates visually meaningful outputs even under the harsh condition of $\epsilon = 0.95$. In contrast, in the single-observation scenario with $K = 1$, the reconstructed images become visually indistinguishable, highlighting the necessity of multiple observations for stable reconstruction under severe noise.

\subsection{Simulation Results in Classification Task}
\begin{figure}[!h]
    \centering
    \includegraphics[width=0.32\textwidth]{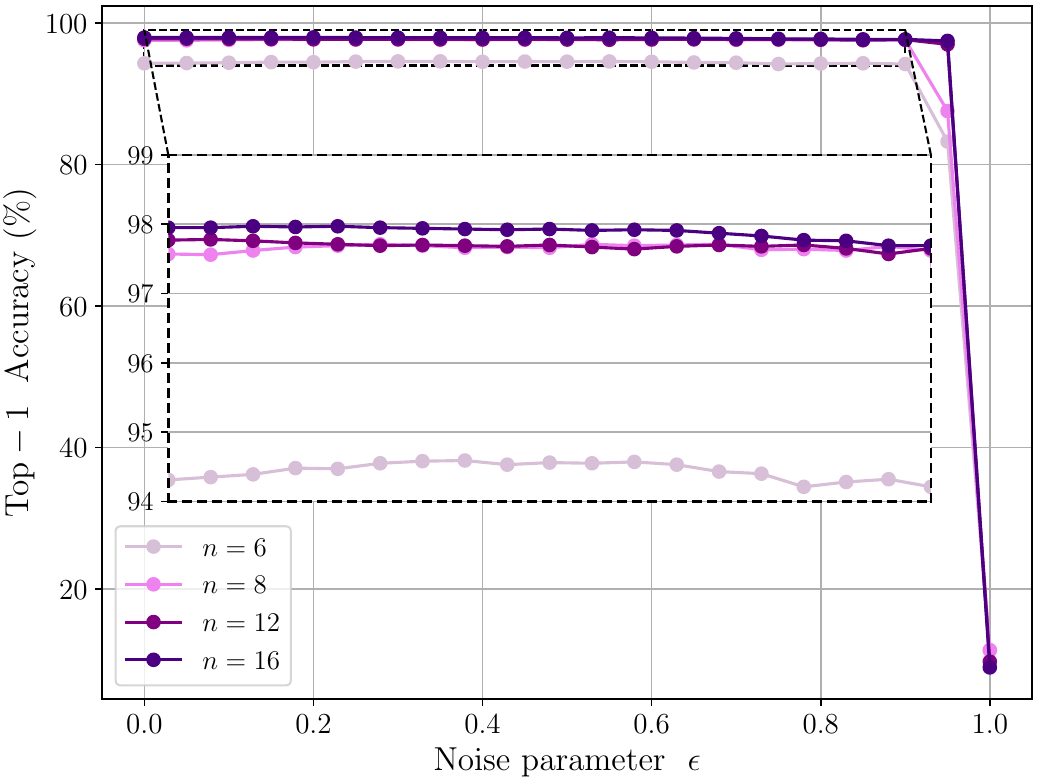}
    \caption{Classification performance comparison against different dimensions of the density matrix.}
    \label{fig:acc_f}
\end{figure}

\begin{figure}[!h]
    \centering
    \includegraphics[width=0.32\textwidth]{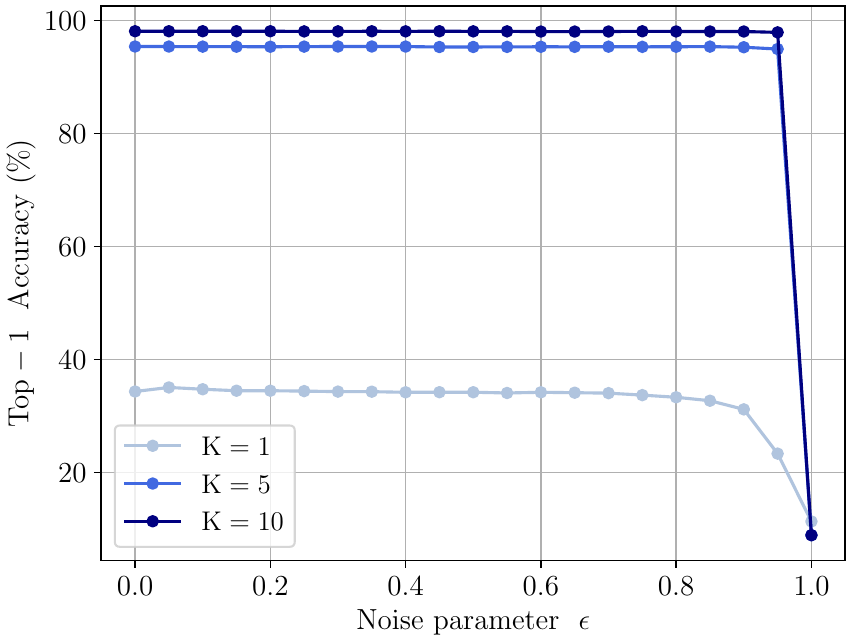}
    \caption{Classification performance comparison against different number of observables.}
    \label{fig:acc_k}
\end{figure}

Fig.~\ref{fig:acc_f} and Fig. \ref{fig:acc_k} present the classification performance with respect to the dimension of the density matrix \(n\) and the number of observations \(K\), respectively. The results show that a smaller density matrix dimension \(n\) leads to consistently better performance across all \(\epsilon\) values, while increasing \(K\) also improves accuracy under all conditions. These trends align with the previously observed results in Fig.~5, further confirming that preserving more information during compression and leveraging multiple observations both contribute to enhanced robustness and classification performance.

\section{Discussion}
\label{sec:discussion}

In this work, we have proposed a novel end-to-end learnable quantum transcoding framework that integrates neural network-based data compression with quantum state encoding via Cholesky decomposition. By aligning the degrees of freedom of the latent representation with those of a density matrix, the proposed method optimizes qubit utilization and improves encoding efficiency over traditional state vector-based approaches.
Our framework incorporates task- and noise-aware design considerations, enabling robust performance under quantum noise while preserving semantic relevance to downstream tasks. Future work includes extending the framework to accommodate a broader range of quantum noise models and validating its effectiveness on more complex and realistic datasets.

\section*{Acknowledgment}

This work is in part supported by the National Research Foundation of Korea (NRF, RS-2024-00451435 (20\%), RS-2024-00413957 (20\%)), Institute of Information \& communications Technology Planning \& Evaluation (IITP, RS-2021-II212068 (10\%), RS-2025-02305453 (15\%), RS-2025-02273157 (15\%), RS-2025-25442149 (10\%) RS-2021-II211343 (10\%)) grant funded by the Ministry of Science and ICT (MSIT), Institute of New Media and Communications (INMAC), and the BK21 FOUR program of the Education, Artificial Intelligence Graduate School Program (Seoul National University), and Research Program for Future ICT Pioneers, Seoul National University in 2025.

\bibliographystyle{IEEEtran}
\bibliography{conference}

\end{document}